\documentclass[twocolumn,showpacs,preprintnumbers,amsmath,amssymb,pre]{revtex4-1}

\usepackage{graphicx}
\usepackage{amsmath}
\usepackage{bm}
\usepackage{color}
\usepackage{dcolumn}
 \usepackage{ulem}

\begin{document}


\newcommand{\udarrow}[2]{\smash{\mathop{%
  \hbox to 0.6cm{$\rightleftharpoons$}}\limits^{#1}\limits_{#2}}}
\newcommand{\p}{\partial}

\newcommand{\bea}{\begin{eqnarray}}
\newcommand{\eea}{\end{eqnarray}}
\newcommand{\be}{\begin{equation}}
\newcommand{\ee}{\end{equation}}
\newcommand{\bse}{\begin{subequations}}
\newcommand{\ese}{\end{subequations}}
\newcommand{\red}[1]{\textcolor{red}{#1}}
\newcommand{\green}[1]{\textcolor{green}{#1}}
\newcommand{\blue}[1]{\textcolor{blue}{#1}}
\newcommand{\comment}[1]{}
\renewcommand{\ss}[1]{_{\hbox{\tiny #1}}}

\title{
Frozen states and order-disorder transition in the dynamics of
confined membranes
}

\author{Thomas Le Goff$^{1}$,  Paolo Politi$^{2,3}$, and Olivier Pierre-Louis$^1$}
\affiliation{
$^1$ Institut Lumi\`ere Mati\`ere, UMR5306 Universit\'e Lyon 1-CNRS, Universit\'e de Lyon 69622 Villeurbanne, France\\
$^2$ Istituto dei Sistemi Complessi, Consiglio Nazionale delle Ricerche, Via Madonna del Piano 10, 50019 Sesto Fiorentino, Italy\\
$^3$ INFN Sezione di Firenze, via G. Sansone 1, 50019 Sesto Fiorentino, Italy.
}

\date{\today}

\begin{abstract}
The adhesion dynamics of a membrane confined between two permeable walls is studied
using a two-dimensional hydrodynamic model.
The membrane morphology decomposes into adhesion patches on the upper and 
the lower walls and obeys a nonlinear evolution equation that resembles
that of phase separation dynamics, which is known to lead to coarsening, i.e. to the endless
growth of the adhesion patches.
However, due to the membrane bending rigidity
the system evolves towards a frozen state without coarsening.
This frozen state exhibits an order-disorder transition
when increasing the permeability of the walls.
\end{abstract}

\maketitle

\section{Introduction}

Two-state continuum models~\cite{Hohenberg1977,Cahn1958}, such as the time-dependent Ginzburg-Landau (TDGL) equation
or the Cahn-Hilliard (CH) equation have been widely studied as a paradygm of phase transition
dynamics in various systems, such as magnetism, liquid-liquid phase separation,
or wetting. These models exhibit a phenomenology characterized
by their coarsening behavior, i.e. the perpetual increase
of the typical lengthscale of the homogeneous zones (where one phase only is present).
In this paper, we propose a one-dimensional two-state continuum model inspired by
adhesion of confined membranes, which gives rise
to a different phenomenology without coarsening and with an order-disorder
transition.

Our motivation is to investigate the adhesion dynamics 
of lipid membranes in biological systems.
Lipid membranes are ubiquitous in living organisms.
They are the main constituent of the cell membrane~\cite{Boal2002},
and also appear in stacks, e.g. in the stratum corneum of the skin~\cite{Bjorklund2013,Das2009a,Das2009b}.
It is therefore crucial to study their physical properties, and especially adhesion,
in order to understand their biological functions.
Adhesion of membranes on substrates~\cite{Perstin2012,Jablin2011,Daillant2005}, 
may include various physical ingredients, such as e.g.
van der Waals attraction and hydration forces~\cite{Swain2001},
ligand-receptor pairs~\cite{Zuckerman1995,Lipowsky1996}, 
interactions with the cytoskeleton~\cite{Speck2012},
osmotic pressures~\cite{Hemmerle2013},
or entropic interactions~\cite{Helfrich1978,Freund2013,Auth2013}.
In this paper, we do not describe these specific ingredients, 
and we rather consider an effective adhesion potential, with a potential minimum
corresponding to an equilibrium adhesion state close to the substrate\cite{Swain2001}.

\begin{figure}
\begin{center}
\includegraphics[height=3.5 cm]{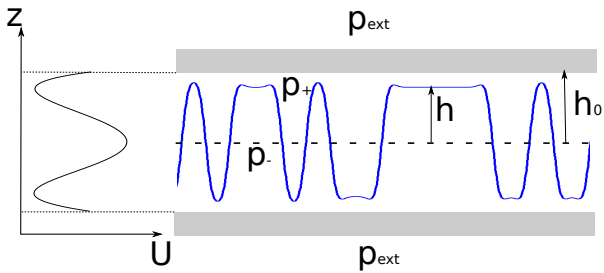}
\end{center}
\caption{
(Color online) 
Schematics of a membrane confined between two permeable walls.
}
\label{fig1}
\end{figure}

The main goal of our work is to study the 
consequences of confinement on the nonlinear dynamics of membrane adhesion. 
In order to mimic confinement within the simplest possible
setting, we consider a membrane located between two parallel flat walls.
The membrane then experiences a total potential which is the sum of the
adhesion potentials of the two substrates.
When the distance between the walls is wider than the
equilibrium distance of a supported membrane on a single wall,
the membrane experiences a double-well potential with a minimum
near each wall, as shown in Fig.~\ref{fig1}.
Such a double-well potential can be found in different instances
in biological systems.
First, in cell adhesion, this double-well potential could account for the possibility of a membrane
to attach to the cytoskeleton inside the cell or to a substrate outside the cell. 
Moreover, in membrane stacks~\cite{Auth2013,Hemmerle2013}, each membrane may adhere
to its neighbors within the stack.
Furthermore, double-well potentials are also found to arise
in the presence of ligands of two different lengths
which enforce two different equilibrium distances in cell-cell adhesion~\cite{Weikl2009}.
In addition, they are also observed experimentally in the combined presence of ligands and van der Waals
attraction which respectively induce  short-range and long-range
attractive potentials, and of glycocalyx and other grafted polymers
which induce a soft repulsion at intermediate scales~\cite{Bruinsma2000,Sengupta2010}.

As a consequence of the double-well, the two walls compete for the adhesion of the membrane,
which is expected to adhere partially on the upper wall,
and partially on the lower wall. 
At first sight, such a decomposition into adhesion patches 
might exhibit some similarity with
phase separation dynamics~\cite{Hohenberg1977},
the membrane height $h(x,t)$ playing the role of the order parameter.
However, in contrast to usual interfaces 
which are controlled by surface tension,
membranes exhibits bending rigidity~\cite{Canham1970,Helfrich1973}:
the membrane energy density is proportional to the mean curvature
squared instead of being proportional to the area.
We shall see in the following that this feature leads to a novel
phenomenology with frozen states: adhesion patches do not grow
and coarsening is absent. From an analysis of the 
nonlinear steady-states, we argue that these
frozen steady-states result from the locking of 
bending-induced membrane oscillations into each-other.

Our results could have some relevance in
a recent debate about the 
formation and stability of finite-size adhesion domains
in cell adhesion.
Different studies have suggested the crucial role of the clustering 
of ligand-receptor pairs~\cite{Lorz2007,ReisterGottfried2008,Bihr2012}, 
of the disorder of the environment~\cite{Speck2012},
of the trapping of ligands in membrane partitions~\cite{Kusumi2005},
or of the active remodeling of the cytoskeleton~\cite{DelanoeAyari2004}. 
We wish to stress that our model where adhesion is driven by
a simple distance-dependent free energy potential
does not account for the full complexity of specific adhesion in cells,
which involves, e.g., the attachement-detachment, diffusion, and interactions of ligand-receptor pairs,
and other ingredients mentioned above.
However, our results indicate a reduced set
of physical ingredients which allows one to obtain finite adhesion patches:
bending rigidity and confinement.

Furthermore, in order to account for the porous character of biological substrates
on which the membrane may adhere,
such as the cytoskeleton, collagen, or endothellial tissues,
we consider walls with arbitrary permeabilities.
Such a tunable permeability is also an important feature
of membrane stacks in the stratum corneum~\cite{Bjorklund2013,Das2009a,Das2009b}.
Our modeling suggests that the spatial organization of the frozen states is controlled
by the permeability of the walls. Indeed, the membrane profile
exhibits a periodic ordered structure for impermeable walls, and becomes disordered
when the wall permeability is increased.
This difference can be traced back to the 
consequences of the permeability on the initial linear instability.

In the following, we start in Section~\ref{s:hydro} with a presentation of the hydrodynamic model, and
we derive a general evolution equation for a membrane between two walls in the lubrication limit.
Then, in Section~\ref{s:cons_non_cons}, we consider the limits of small and large wall permeabilities.
The numerical solution of these limits is discussed in Section~\ref{s:numerical}.
These results are discussed in the light of a linear stability analysis
in Section~\ref{s:lin_stab}, and of an analysis of the nonlinear steady-states
in Section~\ref{s:steady_states}. Finally, we summarize our results
in the last Section.

\section{The hydrodynamic model and the lubrication regime}
\label{s:hydro}

We consider a membrane in a liquid confined between two parallel walls
located in $z=\pm h_0$ (see Fig.~\ref{fig1}).
We focus on the limit of small Reynolds numbers, and the liquid obeys the Stokes
equation:
\be
\nabla p_\pm -\mu\Delta {\mathbf v}_\pm =0 ,
\ee
where the subscript $\pm$ indicates the fluid above $(+)$ or below $(-)$ the membrane at $z=h(x,t)$,
$p_\pm(x,z)$ is the pressure, $\mu$ is the dynamic viscosity, and ${\mathbf v}_\pm= (v_{x\pm },v_{z\pm })$ 
is the liquid velocity. 

Next, we need to define the boundary conditions at the walls and at the membrane, which separates
the upper and lower fluids.
At the walls, the tangential component of the velocity vanishes because we assume no-slip conditions, while
the normal component depends on wall permeability $\nu$:
\bea
v_{x\pm }|_{z=\pm h_0}&=&0 ,
\label{e:bc_wall1}\\
v_{z\pm}|_{z=\pm h_0}&=&\pm \nu (p_\pm-p_{ext})
\label{e:bc_wall2},
\eea
where $p_{ext}$ is a constant pressure outside the walls.

Boundary conditions at membrane are more involved. First, following
Molecular Dynamics simulations on lipid membranes~\cite{Mueller2009,VonHansen2013} we also
assume no-slip at the membrane,
\be
\mathbf v_+|_{z=h(x,t)}=\mathbf v_-|_{z=h(x,t)} .
\label{e:bc_memb1}
\ee
Then, mechanical equilibrium at the membrane imposes
\be
(\mathbf \Sigma_+-\mathbf \Sigma_-) \cdot \mathbf n = \mathbf f ,
\label{e:bc_memb2}
\ee
where $\Sigma_{ij}=\mu(\partial_iv_j+\partial_jv_i)-p\delta_{ij}$ is the stress tensor in the fluid,
$\mathbf n$ is the membrane normal, 
and $\mathbf f=-\delta{\cal E}/\delta\mathbf r$ is the force exerted by the membrane.
This force derives from the energy
\begin{eqnarray}
{\cal E}=\int ds \left[\frac{\kappa}{2}C^2+{\cal U}(h)\right] ,
\label{e:E}
\end{eqnarray}
where $s$ is the arclength along the membrane, 
$C=-\partial_{xx}h/[1+ (\partial_xh)^2]^{3/2}$ is the local membrane curvature, $\kappa$ is the bending rigidity,
and ${\cal U}(h)$ is the double-well adhesion potential, as shown in the schematic in Fig.\ref{fig1}.

Finally, in order to focus on dynamics
within a large contact area and to discard boundary effects, 
we impose periodic boundary conditions along $x$
in a large system of total length $L$.

The main approximation allowing to obtain the evolution equation for the membrane profile
is the small slope approximation $\partial_x h(x,t)\ll 1$, while
the height itself can be finite, i.e of the order of $h_0$.
The main lines of the derivation are reported in Appendix~\ref{a:lubrication}.
Using the standard lubrication expansion~\cite{Oron1997}, we obtain 
\begin{eqnarray}
&&\partial_{t}h=\partial_{x}
\left[-\frac{h_{0}^{3}}{24\mu}\left(1-\frac{h^{2}}{h_{0}^{2}}\right)^{3}\partial_{x}f_{z}
+\frac{3}{4}j\frac{h}{h_{0}}\left(\frac{h^{2}}{3h_{0}^{2}}-1\right)\right]
\nonumber \\
&&\hspace{1 cm}+\frac{\nu}{2}f_{z},
\label{e:model_h}
\end{eqnarray}
where  the membrane force is
\begin{eqnarray}
f_z=-\kappa\partial_{x}^4h-{\cal U}'(h),
\label{e:model_fz}
\end{eqnarray}
and the total liquid flow rate $j$ along $x$,
\begin{eqnarray}
j=\int_{-h_0}^{h}\!\!\!\!\!\!dz\,u_{x-}+\int_{h}^{+h_0}\!\!\!\!\!\!\!\!dz\,u_{x+} ,
\end{eqnarray}
obeys the differential equation
\begin{eqnarray}
-\frac{h_{0}^{3}}{3\mu\nu}\partial_{xx}j+j&=&\frac{1}{2}\frac{h_{0}^{3}}{\mu}\frac{h}{h_{0}}
\left(1-\frac{h^{2}}{3h_{0}^{2}}\right)\partial_xf_z .
\label{e:model_j}
\end{eqnarray}
Two remarks on above equation are in order. First, the equation
is nonlocal in space. This nonlocality is seen from the fact
that $j$ obeys a time-independent differential equation (\ref{e:model_j}).
This constrain comes from the incompressibility of the liquid.
Second, the dynamics is variational, i.e. $\partial_t{\cal E}\leq 0$,
where the energy ${\cal E}$ is given by Eq.~(\ref{e:E}). In the small slope approximation,
the curvature is simply $C=-\partial_{xx}h$ and
\be
{\cal E}=\int dx \left[\frac{\kappa}{2}(\partial_{xx}h)^2+{\cal U}(h)\right] .
\ee

\section{Conserved and non-conserved limits}
\label{s:cons_non_cons}

We are now going to consider
two important limiting cases of Eqs.(\ref{e:model_h}-\ref{e:model_j}), which are better defined
using the reduced wall permeability  
\begin{eqnarray}
\bar\nu=\frac{12\mu\kappa^{1/2}\nu}{h_0^2{\cal U}_0^{1/2}} ,
\end{eqnarray}
where the energy scale ${\cal U}_0$ is such that ${\cal U}(h)={\cal U}_0U(H)$,
where $U(H)$ is of order one.
In the limit of large permeabilities $\bar\nu\rightarrow\infty$, we obtain
\begin{eqnarray}
\partial_TH=-\partial_{X}^4H-U'(H) ,
\qquad \mbox{[TDGL4]}
\label{e:TDGL4}
\end{eqnarray}
where 
$H=h/h_0$,  $X=[{\cal U}_0/(\kappa h_{0}^{2})]^{1/4}x$, 
and $T=t\nu {\cal U}_0/(2h_{0}^{2})$.
In this limit the nonlocality induced by incompressibility vanishes and
the resulting equation has a manifest nonconserved character.
More precisely, Eq.(\ref{e:TDGL4}) bears a strong resemblance to the
standard Time-Dependent Ginzburg-Landau (TDGL) equation, $\partial_TH=\partial_{X}^2H-U'(H)$,
which describes phase separation for a non-conserved order parameter~\cite{Hohenberg1977}. 
However in Eq.(\ref{e:TDGL4}), the linear stabilizing term is 4th order instead of 
being a 2nd order derivative, because
it physically derives from bending rigidity rather than from surface tension.
For this reason, we denote Eq.(\ref{e:TDGL4}) as ``TDGL4".

In the opposite limit of impermeable walls, $\bar\nu=0$, we obtain
\begin{eqnarray}
\partial_TH &=&\partial_X\Bigl\{(1-H^2)^3\partial_X[\partial_{X}^4H+U'(H)]
\nonumber\\
 &&+J H (\frac{H^2}{3}-1)\Bigr\}, \qquad\mbox{[non-local CH4]}
\label{e:CH4}
\\
J&=&-\frac{9}{L}\int_0^L\!\!{\rm d}X\,H\left(1-\frac{H^2}{3}\right) \partial_X[\partial_{X}^4H+U'(H)] ,\quad
\label{e:J_CH4}
\end{eqnarray}
where the time variable now exhibits a different normalization
$T={\cal U}_0^{3/2}t/(24\mu\kappa^{1/2})$,
and $J=18j\mu\kappa^{1/4}/(h_0^{3/2}{\cal U}_0^{5/4})$.
For vanishing permeabilities, the resulting equation is conserved, because the 
(incompressible) fluid remains confined between the walls. 
As a consequence, the membrane evolution equation
shares similarities with the Cahn-Hilliard (CH) equation
$\partial_TH=\partial_{XX}[\partial_{X}^2H-U'(H)]$,
which describes phase separation for a conserved order parameter~\cite{Hohenberg1977,Cahn1958}.
However, there are several differences: 
(i) The 4th-order derivative in the stabilizing term. This difference was expected,
in line with the nonconserved case.
(ii) The membrane mobility $\sim(1-H^2)^3$ vanishes as  $H\rightarrow\pm1$
due to the well known divergence of viscous dissipation
when the membrane approaches the walls~\cite{Oron1997,Happel1983}.
(iii) The non-local effects related to $J$.
The nonlocality is now
manifest in the expression of $J$ as an integral over the whole system in Eq.~(\ref{e:J_CH4}).
In the following, we denote Eq.(\ref{e:CH4}) as the ``non-local CH4" equation.

\section{Numerical study of membrane dynamics}
\label{s:numerical}

As a preamble, before studying the dynamics of Eqs.(\ref{e:TDGL4},\ref{e:CH4})
in extended systems,
we shall recall the well known dynamics arising from the standard TDGL and CH equations:
the profile $H(X,T)\equiv 0$ is unstable and it develops flat regions
where $H$ is approximately equal to the values of one or the other minimum of the 
double-well potential $U(H)$. In the language of our paper, the regions
where the membrane lies in a minimum of the potential correspond to adhesion patches.
The zones separating two flat regions are called kinks.
Within the TDGL or CH models, pairs of kinks collide and annihilate, thereby leading to the decrease
of the number of adhesion patches.
The typical size $\lambda$ of these patches therefore exhibits an endless increases in time.
This process is called coarsening.

In contrast, the numerical solution of TDGL4 and non-local CH4 
does  not exhibit any coarsening. In order to support this statement 
with numerical simulations of the evolution equations,
we need to use an explicit form of the two-well potential $U$.
However, in all other sections above and below, the profile of $U$ is kept arbitrary.
We have chosen the standard quartic potential
\begin{eqnarray}
U_4(h)=-H_m^2\frac{H^2}{2}+\frac{H^4}{4}, 
\end{eqnarray}
which exhibits minimums at $H=\pm H_m$, with $H_m<1$.
In the simulations, we use $H_m=0.9$.

\begin{figure}
\begin{center}
\includegraphics[height=8 cm]{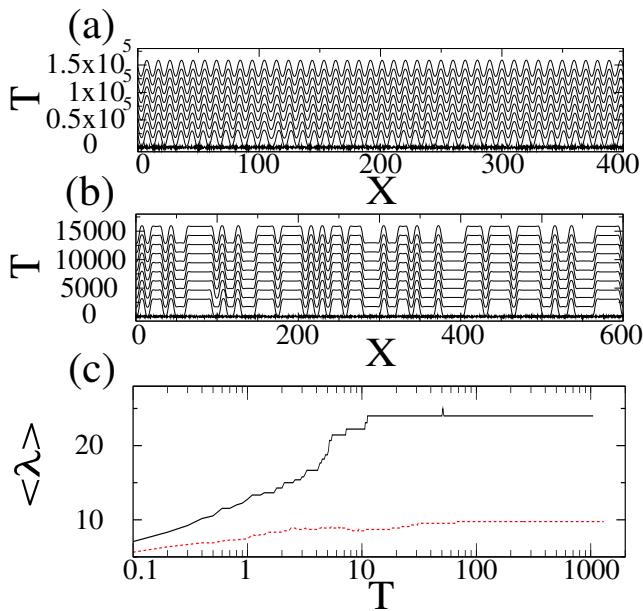}
\end{center}
\caption{
(Color online) 
Arrested dynamics and order-disorder transition.
(a) Non-permeable case. Frozen ordered patterns
obtained from the numerical solution of
the CH4 equation, Eq.(\ref{e:CH4}).
(b) Permeable case. Frozen disordered
patterns obtained from the numerical solution 
of the TDGL4 equation, Eq.(\ref{e:TDGL4}).
In (a,b) the vertical scale is increased by a factor
$\sim10$ for a better visibility of the membrane morhology.}
(c) Saturation of the spatially averaged wavelength $\langle \lambda \rangle$
as a function of time. The black solid line and the 
red dotted line correspond to TDGL4 Eq.(\ref{e:TDGL4})
and CH4 Eq.(\ref{e:CH4}) respectively.
\label{fig2}
\end{figure}

Starting from small random initial conditions we find that
after a short transient the membrane forms a frozen pattern,  
as shown in Fig.\ref{fig2}(a,b). In order to gain quantitative insights
on the evolution of the system, we define the average
wavelength $\langle\lambda\rangle$ as the average
distance between two consecutive points obeying $h=0$ and $\partial_xh>0$.
The plot of  $\langle\lambda\rangle$ as a function of time in Fig.\ref{fig2}(c)
shows a clear saturation after a time of the order of 10 to 30 in reduced
units. Furthermore,
while the frozen pattern is ordered  and periodic
in the presence of impermeable walls, it
is clearly disordered for permeable walls.
We stress that we have observed no difference between the numerical solution
of non-local CH4 and Eq.(\ref{e:CH4}) with $J=0$,
simply denoted as CH4 in the following.

As a first remark on the numerical results, we indicate that
simulations with other forms of the double-well potential $U$
have shown no qualitative difference in the results. 
However, quantitative changes can be observed. As an important
example, when $H_m \rightarrow 1$, the conserved dynamics Eq.~(\ref{e:CH4})
slow down considerably in the late stages because the mobility
term $(1-H^2)^3$ is small in the plateaus between the kinks
where $H$ is close to $H_m$. In constrast, there is no similar effect in 
the non-conserved case Eq.~(\ref{e:TDGL4}).

A second remark:
the final ordered  state obtained in
Fig.~\ref{fig2}(a) for CH4 does not evolve further if used as initial
configuration for TDGL4. And vice versa, the final disordered
state of TDGL4  in Fig.~\ref{fig2}(b) does not evolve under CH4 dynamics: 
this is exactly what we observe from the numerical solution of the equations.
This leads to two important conclusions: (i) the conseved and
non-conserved equation seem to share the same stable steady-states;
(ii) even though distinct ordered and disordered states are
robustly observed with random initial conditions, the final state may also depend
on peculiar initial conditions.

\begin{figure}
\begin{center}
\includegraphics[height=11 cm]{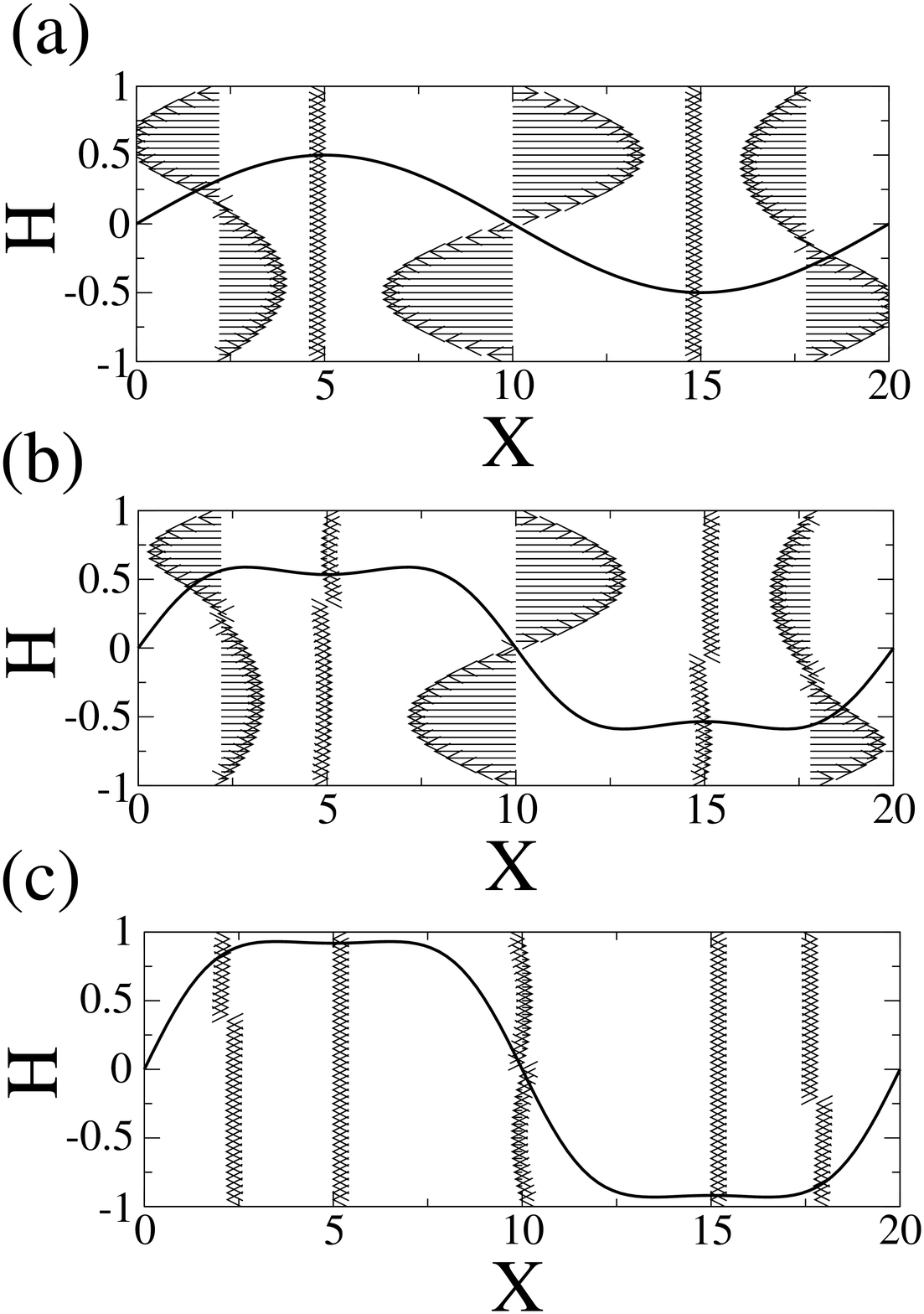}
\end{center}
\caption{
Snapshots of hydrodynamics flows and membrane profile during 
the dynamics in the conserved case Eq.~(\ref{e:CH4}).
Horizontal arrows represent the hydrodynamic flow.
(a) The initial membrane profile is a single period of
a sinusoid. (b) Intermediate times.
(c) The final membrane profile exhibits plateaus
separated by kinks.
}
\label{fig2.5}
\end{figure}

Third remark, once we have the dynamical profile of the membrane,
we also have access to the full hydrodynamic flow
during the evolution of the membrane using Eq.~(\ref{e:poiseuille_lubrication}). As an example,
we show the flow around an initially sinusoidal
membrane profile in the conserved dynamics in Fig.~\ref{fig2.5}.

Finally, we observed that the normalized slopes 
remain finite in all simulations, i.e. $\max|\partial_XH|\sim 1$ at all times. As a consequence,
the small slope approximation $\partial_xh\ll 1$ is self-consistent:
if this assumption is true initially, it remains true for all times.

In the next 
sections, we propose some analytical results which 
confirm the scenario proposed by the numerical solution
of the membrane dynamics.

\section{Linear stability analysis of flat membranes}
\label{s:lin_stab}

As a summary of results so far, Fig.~\ref{fig2} highlights two important features:
(i) absence of coarsening 
and (ii) a frozen state which is disordered for 
$\bar\nu=\infty$ (TDGL4) and ordered for $\bar\nu=0$ (non-local CH4 or CH4).
The latter feature can be traced back to the
different behaviors of the two equations with respect to
small perturbations around the average  height $\bar H$.
Inserting  $H(X,T)=\bar H+\delta H {\rm e}^{i\omega T+iqX}$
with $\delta H\ll 1$ in Eq.(\ref{e:TDGL4})
we obtain to linear order the dispersion relation for TDGL4
\be
i\omega=-U''(\bar H)-q^4 . \qquad \mbox{[TDGL4]}
\ee 
As a remark, in the limit of permeable walls and when $U'(\bar H)\neq 0$,
the average height $\bar H$ depends on time. Hence, strictly speaking
the dynamical evolution of the Fourier modes
is not exponential. However, the dispersion relation still provides
a qualitative description of the unstable modes at short times
for $\bar H\neq 0$. In addition,
the linear stability analysis also provides a strictly valid description 
for the case $\bar H=0$ studied in the numerical simulations above,
because $\bar H$ is constant in this case.

In contrast, $\bar H$ is always constant in the conserved equations,
and the exponential time-dependence of the perturbation
amplitude is strictly valid in this case.
The linear dispersion relation for non-local CH4 or CH4 (i.e. with 
or without the $J$ term) provides the same dispersion relation
\be
i\omega= (1-\bar H^2)^3q^2[-U''(\bar H)-q^4] . \qquad \mbox{[CH4]}
\ee

\begin{figure}
\begin{center}
\includegraphics[width=8.5 cm]{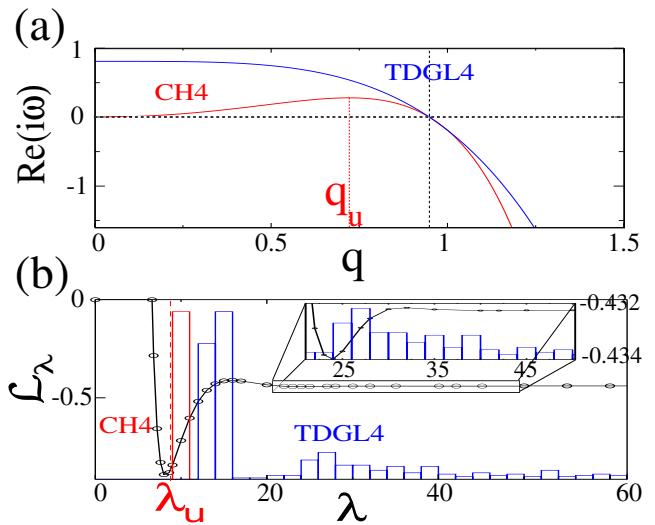}
\end{center}
\caption{
(Color online) 
(a) Linear dispersion relation.
(b) Histogram of the distances between kinks.
The solid line reports the value of minus the total curvature
energy in one steady-state period ${\cal L}_\lambda=-\int_0^\lambda(\partial_{XX}H_{\lambda}(X))^2$,
obtained numerically from the periodic double-kink
solution shown in Fig.~\ref{fig4}.
As discussed in Sec.~\ref{s:steady_states}, ${\cal L}_\lambda$ controls the stability of
the steady-states. The dashed line corresponds to the approximate 
expression of Eq.(\ref{e:L_approx}) with ${\cal L}_0=-0.43225$.
}
\label{fig3}
\end{figure}

Both for permeable and impermeable walls,
an instability, indicated by $i\omega>0$, appears at long wavelength when $U''(\bar H)<0$.
As seen in Fig.~\ref{fig3}(a), while TDGL4 destabilizes all long wavelength modes
with the same growth rate $i\omega\sim -U''(\bar H)$ at $q\rightarrow 0$,
CH4 exhibits a special mode at $q_u=[-U''(\bar H)/3]^{1/4}$ for which the growth rate
is maximum.
Hence, we expect initially a disordered pattern with many wavelengths in the limit of permeable walls,
and an ordered pattern with a single wavelength $\lambda_u=2\pi/q_u$ in the limit of impermeable walls.
In Fig.~\ref{fig3}(b), we have plotted the histogram of $\lambda $, 
the double of the distances between the zeros of $H$
in the frozen state when starting from random initial conditions.
The quantity $\lambda $ is a measure of the local wavelength.
For non-local (and local) CH4, the linear instability produces
an initial periodic pattern with a single wavelength $\lambda_u\approx 2\pi/q_u$,
while for TDGL4, we indeed obtain a wide distribution of distances.

\section{Stability of periodic steady-states}
\label{s:steady_states}

Although linear analysis indicates when we should expect order
or disorder, it does not provide insights about why the
dynamics should freeze, as observed in the simulations.
In order to gain insights on this subject
we study the stability of fully nonlinear periodic steady-states.
The steady-states of the  TDGL and CH equations, solutions of $\partial_{X}^2H-U'(H)=0$,
are known to be periodic with a single maximum in each period.
For each wavelength $\lambda$, there is a unique steady-state.
For Eq.(\ref{e:model_h}), and all its special limits TDGL4, CH4, and non-local CH4, 
the steady-states obey
\begin{eqnarray}
\partial_{X}^4H+U'(H)=0.
\label{e:steady_states}
\end{eqnarray}
It is actually known that Eq.(\ref{e:steady_states})
exhibits not only periodic solutions with several maximums
per period, but also an infinite number
of non-periodic solutions (chaotic along $x$)~\cite{Peletier2001}.
However, we shall show in the following that 
the study of periodic steady-states provides a reasonable 
description of the nonlinear dynamics.

For this purpose, consider a family of periodic steady-states $H_{\lambda}$
parametrized by the wavelength $\lambda$.
We wish to study the stability of a uniform periodic
steady-state under long-wavelength variations of $\lambda$.
Defining a macroscopic variable $\tilde X$ at scales much larger than $\lambda$,
the total energy may be approximated as the integral on the slow variable $\tilde X$
of the energy density in one period
\begin{eqnarray}
 {\cal E}= \int \frac{d\tilde X}{\lambda(\tilde X)}\int_0^{\lambda(\tilde X)}\hspace{-0.5 cm}dX
e_{\lambda(\tilde X)}(X)
\end{eqnarray}
where 
\begin{eqnarray}
e_{\lambda(\tilde X)}(X)=[\partial_{XX}H_{\lambda(\tilde X)}(X)]^2/2+U(H_{\lambda(\tilde X)}(X))
\end{eqnarray}
is the local energy density.
We then consider a small perturbation around the average wavelength $\lambda(\tilde X)=\bar\lambda +\delta\lambda(\tilde X)$.
Since $\delta\lambda(\tilde X)$ is small, the 
total number ${\cal N}=\int d\tilde X/\lambda(\tilde X)$
of steady-state periods in the system is constant, i.e. $\delta {\cal N}=0$, leading to the relation 
\begin{eqnarray}
\lambda \int d\tilde X \delta\lambda(\tilde X) \approx \int d\tilde X \delta\lambda(\tilde X)^2+O(\delta\lambda(\tilde X)^3).
\end{eqnarray}
Using this relation and Eq.(\ref{e:steady_states}), one may then calculate the variation of total energy 
\begin{eqnarray}
\delta {\cal E}
= \frac{\partial_{\bar\lambda}{\cal L}_{\bar\lambda}}{\bar\lambda^2} 
\int d\tilde X [\delta\lambda(\tilde X)]^2+O([\delta\lambda(\tilde X)]^3),
\end{eqnarray}
where
\begin{eqnarray}
{\cal L}_\lambda=-\int_0^\lambda(\partial_{XX}H_{\lambda}(X))^2.
\label{e:L}
\end{eqnarray}
Since we know that the dynamics
always 
decreases ${\cal E}$, 
i.e. $\partial_t{\cal E}\leq 0$,
the perturbation amplitude $\int d\tilde X [\delta\lambda(\tilde X)]^2$
must decrease if $\partial_\lambda{\cal L}_{\lambda}>0$,
and must increase if $\partial_\lambda{\cal L}_{\lambda}<0$.
Hence, the periodic steady-state of wavelength $\bar\lambda$
is stable if $\partial_\lambda{\cal L}_{\lambda}>0$
and unstable if $\partial_\lambda{\cal L}_{\lambda}<0$.
This criterion shows that the stability depends only on the
energy ${\cal E}$, and is independent of the precise
kinetics. This criterion based on the energy is valid for the general Eq.(\ref{e:model_h}),
and its various specific limits (TDGL4, non-local CH4, or CH4).

\begin{figure}
\begin{center}
\includegraphics[height=7 cm]{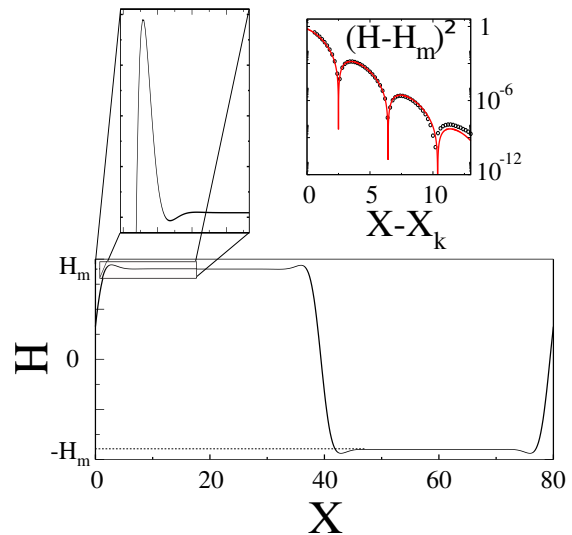}
\end{center}
\caption{
(Color online) 
Periodic  double-kink steady-state profile.
The insets show a zoom on an oscillatory kink tail,
and the oscillations of $(H-H_m)^2$ in log scale away from a kink.
The transient dynamics leading to this 
periodic steady-state is shown in Fig.~\ref{fig2.5}.
}
\label{fig4}
\end{figure}

We use a branch
of steady-state solutions which provide the double-kink solution
shown in Fig.~\ref{fig4} at long wavelengths to calculate ${\cal L}_\lambda$.
Hereafter, we define a kink as
a localized region of the membrane profile going from $\mp H_m$ for $x\rightarrow -\infty$ to $\pm H_m$
for  $x\rightarrow +\infty$.
This branch can for example be obtained 
from the relaxation with TDGL4 of an initial condition
composed of a double kink with $\tanh$ profiles.
In Fig.~\ref{fig3}(b), we have plotted ${\cal L}_{\lambda}$ from this steady-state branch.
We see that $\partial_\lambda{\cal L}_{\lambda}>0$
for the most unstable wavelength of the CH4 or non-local CH4 equations, $\lambda=\lambda_u$.
Hence, our stability criterion explains that the periodic
steady-state reached by the dynamics via the linear
instabilty of CH4 or non-local CH4 is frozen.

The case of the TDGL4 equation is more delicate
to analyze because we start with a disordered state
as discussed earlier. 
However, we see peaks in the histogram of Fig.~\ref{fig3}(b)
in the stable regions with $\partial_\lambda{\cal L}_{\lambda}>0$,
and valleys when $\partial_\lambda{\cal L}_{\lambda}<0$.
This is in agreement with a scenario where
pairs of zeros separated by a distance corresponding to $\partial_\lambda{\cal L}_{\lambda}<0$
are unstable, and the whole system finally recombines
into a configuration where the distance between the zeros
are in the stable regions.
However,
note that for large distances, 
no reorganization is obtained within the simulation time.

A striking feature of the stability criterion in Fig.~\ref{fig3}(b)
is its oscillatory character.
These oscillations originate in the fourth order derivative in Eq.(\ref{e:steady_states}), 
which induces an oscillatory membrane profile 
in the vicinity of the kinks, as shown in Fig.~\ref{fig4}.
Expanding Eq.(\ref{e:steady_states}) in the vicinity
of the minima of potential wells
at $H=H_m$ for $X>X_k$, where $X_k$ is the position of the kink, we find
an explicit expression for the kink tails $H(X)=H_m+R(X-X_k)$, with
\begin{eqnarray}
R(\ell)=A\cos\left(\frac{\ell{U''_m}^{1/4}}{2^{1/2}}+\alpha\right)\exp\left[-\frac{\ell{U''_m}^{1/4}}{2^{1/2}}\right].
\label{e:R(ell)}
\end{eqnarray}
Here $U''_m=U''(H_m)$, and $A>0$ and $\alpha$ are constants depending 
on details of the potential profile.
Since we do not have an analytical expression for the full kink profile,
the exact values of $A$ and $\alpha$ are unknown
and depend on the precise profile of $U$.
 However, a simple argument provides an approximate 
value. Indeed, assuming that the profile $H(X)=H_m+R(X-X_k)$
with $R$ given in Eq.(\ref{e:R(ell)}) extends
beyond its domain of validity up to the center of the kink where $X\rightarrow X_k$,
we request the continuity  of $H$ at $X=X_k$ up to the third derivative,
leading to $H(X_k)=0$, and $\partial_{XX}H(X_k)=0$. As a consequence
of these assumptions, one finds $A=H_m$, and $\alpha=\pi$. 
For the specific case of the quartic potential ${\cal U}(H)={\cal U}_4(H)$
with $H_m=0.9$, these constants can be determined numerically
by fitting the profile of the tail of an isolated kink
with Eq.(\ref{e:R(ell)}), as shown in the inset of Fig.~\ref{fig4}.
We then find values which are close to the approximate predictions:
$A=0.87$, and $\alpha=2.72$.

For large distances between the kinks $\lambda\gg 1$, the 
behavior of ${\cal L}_\lambda$ is actually dominated by the asymptotic tails
of the kinks, and substituting Eq.(\ref{e:R(ell)}) into Eq.(\ref{e:L}), we find to leading order:
\begin{eqnarray}
&&{\cal L}_\lambda\approx {\cal L}_0+
\nonumber \\
&&A^2{U''_m}\lambda\cos\left(\frac{\lambda{U''_m}^{1/4}}{2^{3/2}}+2\alpha\right)
\exp\left[-\frac{\lambda{U''_m}^{1/4}}{2^{3/2}}\right] ,
\label{e:L_approx}
\end{eqnarray}
where ${\cal L}_0$ is an unknown constant. This expression is in 
good agreement with the value of ${\cal L}$ obtained from the 
numerical profile of the steady-state branch, as shown in Fig.~\ref{fig3}(b).
As discussed above, the stability criterion is related to the sign of
\begin{eqnarray}
\partial_\lambda{\cal L}_\lambda&\approx& -\frac{1}{2}A^2{U''_m}^{5/4}\lambda\cos\left(\frac{\lambda{U''_m}^{1/4}}{2^{3/2}}+2\alpha-\frac{\pi}{4}\right)
\nonumber \\
&&\times \exp\left[-\frac{\lambda{U''_m}^{1/4}}{2^{3/2}}\right] .
\end{eqnarray}
This expression shows explicitly the oscillatory character of the stability
as a function of the distance between kinks.

\section{Final considerations}

As a summary, we have derived a nonlinear and nonlocal dynamical equation, see Eq.~(\ref{e:model_h}), 
from a hydrodynamic model for a membrane separating two
incompressible fluids and confined between two rigid walls, see Fig.~\ref{fig1}.
This equation has been studied numerically and analytically in the limit of large
wall permeability ($\bar\nu\to\infty$), leading to the nonconserved Eq.~(\ref{e:TDGL4}), and in the limit
of vanishing wall permeability ($\bar\nu\to 0$), leading to the conserved Eq.~(\ref{e:CH4}).

The bending rigidity of the membrane induces a novel class of behavior.
Indeed both  for small and large $\bar\nu$, the system evolves towards a frozen
state, the details of which depend on the initial state. Generic, random initial configurations lead to a
disordered state for large $\bar\nu$ (conserved case) and to an ordered periodic state for vanishing 
$\bar\nu$ (nonconserved case).
The non-local character of the dynamics appears to be either vanishing ($\bar\nu\to\infty$)
or irrelevant ($\bar\nu\to 0$).

The orders of magnitude
of the lengthscales and time-scales of the patterns discussed in this paper
should be observable experimentally.
Indeed, following Ref.~\cite{Swain2001},
we consider as an example an attractive Van der Waals interaction 
and the hydration repulsion between a membrane
and a substrate. 
Using a gap $2h_0=20$nm with  $\bar{h}=0$,
the most unstable wavelength in the case of impermeable
walls (non-local CH4) is $\lambda_{u}=2\pi/q_u\approx 350$nm 
and $t_{u}\approx1\times10^{-2}s$.

Besides the need of generalization
of our approach to two-dimensional membranes,
one important perspective of our work
is to test the robustness of the frozen
states with respect to various additional physical
ingredients. As an example, a membrane tension $\sigma$
can be added to the model, leading to an additional 
stabilizing term $\sigma\partial_{xx}h$ in the expression of
the membrane force Eq.(\ref{e:model_fz}).
For large enough tensions, the oscillations in 
the kink tails disappear. As expected, the dynamics for large
tensions is similar to that of TDGL or CH, with logarithmic
coarsening. From a simple dimensional analysis, this behavior is expected for tensions
larger than  $\sigma_c\sim ({\cal U}_0\kappa)^{1/2}/h_0$,
with a prefactor of the order of 1.
A detailed account of this transition confirms this prediction, and will
be provided elsewhere \cite{LeGoff2014}. 
Using once again numbers from Ref.\cite{Swain2001},
we find $\sigma_c\sim 10^{-2}$J.m$^{-2}$. Values for the tension 
of supported membranes extracted from experiments are 
in the range $\sigma\approx 10^{-5} - 10^{-3}$J 
from Refs.~\onlinecite{Swain2001,Malaquin2010}. 
As a consequence the tensions observed in supported membranes are 
much smaller than $\sigma_c$, and their effects should be negligible.
However, the area increase (or decrease) 
in the kinks during the formation (or annihilation) of
adhesion patches could also lead to additional tension
effects.

Other ingredients, such as potential asymmetry and noise could also
destabilize the frozen states reported here. We plan
to report along these lines in the near future.

\begin{acknowledgments}
We acknowledge support from Biolub grant ANR-12-BS04-0008 (TLG,PP,OPL), and from INFN (OPL).
\end{acknowledgments}

\begin{appendix}
\section{Lubrication limit}
\label{a:lubrication}

Here we provide the main lines of the derivation
of an evolution equation for the membrane in the lubrication limit.
We start with a slightly more general description as compared to 
the one discussed in the main text. Indeed, we describe the hydrodyanmics
with the full Navier-Stokes equations, including inertial effects.
Consider a fluid in two dimensions $x,z$:
\begin{eqnarray}
\rho(\partial_{t}v_{x}+v_{x}\partial_{x}v_{x}+v_{z}\partial_{z}v_{x})&=&-\partial_{x}p+\mu\nabla^{2}v_{x},\nonumber
\\
\rho(\partial_{t}v_{z}+v_{x}\partial_{x}v_{z}+v_{z}\partial_{z}v_{z})&=&-\partial_{z}p+\mu\nabla^{2}v_{z},\nonumber
\end{eqnarray}
where $\rho$ is the density of the fluid, 
and the other notations are defined in the main text.

We define a small parameter $\epsilon=h_{0}/{\ell}\ll1$ where $\ell$ 
is the typical extent of the adhesion patches along $x$.
We may then define dimensionless variables $X={\epsilon x}/{h_{0}}$, $Z={z}/{h_{0}}$.
Following the usual procedure for the lubrication expansion~\cite{Oron1997},
we also use normalized velocities $V_{X}={v_{x}}/{v_{0}}$ and $V_{Z}={v_{z}}/(\epsilon v_{0})$,
and pressure $P= {\epsilon h_{0}}/(\mu v_{0})$, where $v_0$ is the typical fluid velocity.
With these new variables, we obtain
\begin{eqnarray}
&&\epsilon \operatorname{Re}(\partial_{T}V_{X}+V_{X}\partial_{X}V_{X}+V_{Z}\partial_{Z}V_{X})=
\nonumber \\
&&-\partial_{X}P+\partial_{Z}^{2}V_{X}+\epsilon^{2}\partial_{X}^{2}V_{X},\nonumber
\\
&&\epsilon^{3}\operatorname{Re}(\partial_{T}V_{Z}+V_{X}\partial_{X}V_{Z}+V_{Z}\partial_{Z}V_{Z})=
\nonumber \\
&&-\partial_{Z}P+\epsilon^{2}(\partial_{Z}^{2}V_{Z}+\epsilon^{2}\partial_{X}^{2}V_{Z}),\nonumber
\end{eqnarray}
where $\operatorname{Re}={\rho v_{0}h_{0}}/{\mu}$ is the Reynolds number. 
Assuming that Re is at most of order one, and in
the lubrication approximation $\epsilon\rightarrow 0$, 
we obtain to leading order $-\partial_{X}P+\partial_{Z}^{2}V_{X}=0$,
and $-\partial_{Z}P=0$. As a consequence $P$ depends only
on $X$, and $V_X$ exhibits a simple quadratic form 
\begin{equation}
V_X=\frac{Z^2}{2}\partial_{X}P+AZ+B,
\label{e:poiseuille_lubrication}
\end{equation}
where $P$, $A$ and $B$ are $3$ unknown functions of
$X$ which do not depend on $Z$. Since the fluid may have different
velocity profiles above and below the membrane, we obtain
$6$ unknown functions of $X$. It is convenient to define the total flow rate $J$ obeying
\begin{eqnarray}
J=\int_{-1}^{1} dZ\; V_X
\label{e:J_def}
\end{eqnarray}
 as a seventh unknown function of $X$.

These seven unknown functions of $X$ are obtained using the
boundary conditions at the wall and at the membrane.
The no-slip conditions at the walls and at the membrane, 
Eqs.(\ref{e:bc_wall1},\ref{e:bc_memb1}), provide three equations. 
Then, mechanical equilibrium at the membrane, Eq.(\ref{e:bc_memb2}), 
leads to two additional equations.
Hence, we have five equations:
\begin{eqnarray}
\strut   V_{X+}\vert_{Z=1}&=&0,       \label{e:bc1}\\
V_{X-}\vert_{Z=-1}&=&0,               \label{e:bc2}\\
V_{X+}\vert_{Z=H}&=&V_{X-}\vert_{Z=H},\label{e:bc3}\\
P_{+}-P_{-}&=&F_{Z},                  \label{e:bc4}\\
\partial_{Z}V_{X+}\vert_{Z=H}&=&\partial_{Z}V_{X-}\vert_{Z=H}. \label{e:bc5}
\end{eqnarray}

Mass conservation and the wall permeability condition, Eq.(\ref{e:bc_wall2}), provide two 
other equations:
\begin{eqnarray}
\partial_XJ&=& -\tilde\nu (P_++P_--2P_{ext}),
\label{e:BCJ}
\\
\partial_{T}H&=&-\frac{1}{2}\partial_{X}(J_{-}-J_{+})+\frac{\tilde\nu}{2}(P_{+}-P_{-}),
\label{e:BCH}
\end{eqnarray}
where $\tilde \nu=\nu\mu\ell^2h_0^{-3}$, and the upper and lower liquid flow rates are defined as
\begin{eqnarray}
J_- &=& \int_{-1}^{H} dZ\; V_X,
\\
J_+ &=& \int_{H}^{1} dZ \; V_X.
\end{eqnarray}
Using the seven equations (\ref{e:J_def}-\ref{e:BCJ})
provides the seven unknowns. Inserting these expressions in Eq.(\ref{e:BCH})
and going back to physical variables
leads to the evolution equation of the membrane, Eq.(\ref{e:model_h}).

\end{appendix}


\begin{thebibliography}{34}%
\makeatletter
\providecommand \@ifxundefined [1]{%
 \@ifx{#1\undefined}
}%
\providecommand \@ifnum [1]{%
 \ifnum #1\expandafter \@firstoftwo
 \else \expandafter \@secondoftwo
 \fi
}%
\providecommand \@ifx [1]{%
 \ifx #1\expandafter \@firstoftwo
 \else \expandafter \@secondoftwo
 \fi
}%
\providecommand \natexlab [1]{#1}%
\providecommand \enquote  [1]{``#1''}%
\providecommand \bibnamefont  [1]{#1}%
\providecommand \bibfnamefont [1]{#1}%
\providecommand \citenamefont [1]{#1}%
\providecommand \href@noop [0]{\@secondoftwo}%
\providecommand \href [0]{\begingroup \@sanitize@url \@href}%
\providecommand \@href[1]{\@@startlink{#1}\@@href}%
\providecommand \@@href[1]{\endgroup#1\@@endlink}%
\providecommand \@sanitize@url [0]{\catcode `\\12\catcode `\$12\catcode
  `\&12\catcode `\#12\catcode `\^12\catcode `\_12\catcode `\%12\relax}%
\providecommand \@@startlink[1]{}%
\providecommand \@@endlink[0]{}%
\providecommand \url  [0]{\begingroup\@sanitize@url \@url }%
\providecommand \@url [1]{\endgroup\@href {#1}{\urlprefix }}%
\providecommand \urlprefix  [0]{URL }%
\providecommand \Eprint [0]{\href }%
\providecommand \doibase [0]{http://dx.doi.org/}%
\providecommand \selectlanguage [0]{\@gobble}%
\providecommand \bibinfo  [0]{\@secondoftwo}%
\providecommand \bibfield  [0]{\@secondoftwo}%
\providecommand \translation [1]{[#1]}%
\providecommand \BibitemOpen [0]{}%
\providecommand \bibitemStop [0]{}%
\providecommand \bibitemNoStop [0]{.\EOS\space}%
\providecommand \EOS [0]{\spacefactor3000\relax}%
\providecommand \BibitemShut  [1]{\csname bibitem#1\endcsname}%
\let\auto@bib@innerbib\@empty
\bibitem [{\citenamefont {Hohenberg}\ and\ \citenamefont
  {Halperin}(1977)}]{Hohenberg1977}%
  \BibitemOpen
  \bibfield  {author} {\bibinfo {author} {\bibfnamefont {P.~C.}\ \bibnamefont
  {Hohenberg}}\ and\ \bibinfo {author} {\bibfnamefont {B.~I.}\ \bibnamefont
  {Halperin}},\ }\href {\doibase 10.1103/RevModPhys.49.435} {\bibfield
  {journal} {\bibinfo  {journal} {Rev. Mod. Phys.}\ }\textbf {\bibinfo {volume}
  {49}},\ \bibinfo {pages} {435} (\bibinfo {year} {1977})}\BibitemShut
  {NoStop}%
\bibitem [{\citenamefont {Cahn}\ and\ \citenamefont
  {Hilliard}(1958)}]{Cahn1958}%
  \BibitemOpen
  \bibfield  {author} {\bibinfo {author} {\bibfnamefont {J.~W.}\ \bibnamefont
  {Cahn}}\ and\ \bibinfo {author} {\bibfnamefont {J.~E.}\ \bibnamefont
  {Hilliard}},\ }\href@noop {} {\bibfield  {journal} {\bibinfo  {journal} {The
  Journal of Chemical Physics}\ }\textbf {\bibinfo {volume} {28}},\ \bibinfo
  {pages} {258} (\bibinfo {year} {1958})}\BibitemShut {NoStop}%
\bibitem [{\citenamefont {Boal}(2002)}]{Boal2002}%
  \BibitemOpen
  \bibfield  {author} {\bibinfo {author} {\bibfnamefont {D.}~\bibnamefont
  {Boal}},\ }\href@noop {} {\emph {\bibinfo {title} {Mechanics of the cell}}}\
  (\bibinfo  {publisher} {Cambridge University Press},\ \bibinfo {year}
  {2002})\BibitemShut {NoStop}%
\bibitem [{Bjo(2013)}]{Bjorklund2013}%
  \BibitemOpen
  \href {\doibase http://dx.doi.org/10.1016/j.ejps.2013.04.022} {\bibfield
  {journal} {\bibinfo  {journal} {European Journal of Pharmaceutical Sciences}\
  }\textbf {\bibinfo {volume} {50}},\ \bibinfo {pages} {638 } (\bibinfo {year}
  {2013})},\ \bibinfo {note} {(Trans)dermal drug delivery: Emerging trends to
  study and overcome the skin barrier}\BibitemShut {NoStop}%
\bibitem [{\citenamefont {Das}\ \emph {et~al.}(2009{\natexlab{a}})\citenamefont
  {Das}, \citenamefont {Olmsted},\ and\ \citenamefont {Noro}}]{Das2009a}%
  \BibitemOpen
  \bibfield  {author} {\bibinfo {author} {\bibfnamefont {C.}~\bibnamefont
  {Das}}, \bibinfo {author} {\bibfnamefont {P.~D.}\ \bibnamefont {Olmsted}}, \
  and\ \bibinfo {author} {\bibfnamefont {M.~G.}\ \bibnamefont {Noro}},\ }\href
  {\doibase 10.1039/B911257J} {\bibfield  {journal} {\bibinfo  {journal} {Soft
  Matter}\ }\textbf {\bibinfo {volume} {5}},\ \bibinfo {pages} {4549} (\bibinfo
  {year} {2009}{\natexlab{a}})}\BibitemShut {NoStop}%
\bibitem [{\citenamefont {Das}\ \emph {et~al.}(2009{\natexlab{b}})\citenamefont
  {Das}, \citenamefont {Noro},\ and\ \citenamefont {Olmsted}}]{Das2009b}%
  \BibitemOpen
  \bibfield  {author} {\bibinfo {author} {\bibfnamefont {C.}~\bibnamefont
  {Das}}, \bibinfo {author} {\bibfnamefont {M.~G.}\ \bibnamefont {Noro}}, \
  and\ \bibinfo {author} {\bibfnamefont {P.~D.}\ \bibnamefont {Olmsted}},\
  }\href@noop {} {\bibfield  {journal} {\bibinfo  {journal} {Biophys. Journal}\
  }\textbf {\bibinfo {volume} {97}},\ \bibinfo {pages} {1951} (\bibinfo {year}
  {2009}{\natexlab{b}})}\BibitemShut {NoStop}%
\bibitem [{\citenamefont {Das}\ \emph {et~al.}(2012)\citenamefont {Das},
  \citenamefont {Noro},\ and\ \citenamefont {Olmsted}}]{Perstin2012}%
  \BibitemOpen
  \bibfield  {author} {\bibinfo {author} {\bibfnamefont {C.}~\bibnamefont
  {Das}}, \bibinfo {author} {\bibfnamefont {M.~G.}\ \bibnamefont {Noro}}, \
  and\ \bibinfo {author} {\bibfnamefont {P.~D.}\ \bibnamefont {Olmsted}},\
  }\href@noop {} {\bibfield  {journal} {\bibinfo  {journal} {Biointerphases}\
  }\textbf {\bibinfo {volume} {7}},\ \bibinfo {pages} {57} (\bibinfo {year}
  {2012})}\BibitemShut {NoStop}%
\bibitem [{\citenamefont {Jablin}\ \emph {et~al.}(2011)\citenamefont {Jablin},
  \citenamefont {Zhernenkov}, \citenamefont {Toperverg}, \citenamefont {Dubey},
  \citenamefont {Smith}, \citenamefont {Vidyasagar}, \citenamefont {Toomey},
  \citenamefont {Hurd},\ and\ \citenamefont {Majewski}}]{Jablin2011}%
  \BibitemOpen
  \bibfield  {author} {\bibinfo {author} {\bibfnamefont {M.~S.}\ \bibnamefont
  {Jablin}}, \bibinfo {author} {\bibfnamefont {M.}~\bibnamefont {Zhernenkov}},
  \bibinfo {author} {\bibfnamefont {B.~P.}\ \bibnamefont {Toperverg}}, \bibinfo
  {author} {\bibfnamefont {M.}~\bibnamefont {Dubey}}, \bibinfo {author}
  {\bibfnamefont {H.~L.}\ \bibnamefont {Smith}}, \bibinfo {author}
  {\bibfnamefont {A.}~\bibnamefont {Vidyasagar}}, \bibinfo {author}
  {\bibfnamefont {R.}~\bibnamefont {Toomey}}, \bibinfo {author} {\bibfnamefont
  {A.~J.}\ \bibnamefont {Hurd}}, \ and\ \bibinfo {author} {\bibfnamefont
  {J.}~\bibnamefont {Majewski}},\ }\href {\doibase
  10.1103/PhysRevLett.106.138101} {\bibfield  {journal} {\bibinfo  {journal}
  {Phys. Rev. Lett.}\ }\textbf {\bibinfo {volume} {106}},\ \bibinfo {pages}
  {138101} (\bibinfo {year} {2011})}\BibitemShut {NoStop}%
\bibitem [{\citenamefont {Daillant}\ \emph {et~al.}(2005)\citenamefont
  {Daillant}, \citenamefont {Bellet-Amalric}, \citenamefont {Braslau},
  \citenamefont {Charitat}, \citenamefont {Fragneto}, \citenamefont {Graner},
  \citenamefont {Mora}, \citenamefont {Rieutord},\ and\ \citenamefont
  {Stidder}}]{Daillant2005}%
  \BibitemOpen
  \bibfield  {author} {\bibinfo {author} {\bibfnamefont {J.}~\bibnamefont
  {Daillant}}, \bibinfo {author} {\bibfnamefont {E.}~\bibnamefont
  {Bellet-Amalric}}, \bibinfo {author} {\bibfnamefont {A.}~\bibnamefont
  {Braslau}}, \bibinfo {author} {\bibfnamefont {T.}~\bibnamefont {Charitat}},
  \bibinfo {author} {\bibfnamefont {G.}~\bibnamefont {Fragneto}}, \bibinfo
  {author} {\bibfnamefont {F.}~\bibnamefont {Graner}}, \bibinfo {author}
  {\bibfnamefont {S.}~\bibnamefont {Mora}}, \bibinfo {author} {\bibfnamefont
  {F.}~\bibnamefont {Rieutord}}, \ and\ \bibinfo {author} {\bibfnamefont
  {B.}~\bibnamefont {Stidder}},\ }\href {\doibase 10.1073/pnas.0504588102}
  {\bibfield  {journal} {\bibinfo  {journal} {Proceedings of the National
  Academy of Sciences of the United States of America}\ }\textbf {\bibinfo
  {volume} {102}},\ \bibinfo {pages} {11639} (\bibinfo {year}
  {2005})}\BibitemShut {NoStop}%
\bibitem [{\citenamefont {Swain}\ and\ \citenamefont
  {Andelman}(2001)}]{Swain2001}%
  \BibitemOpen
  \bibfield  {author} {\bibinfo {author} {\bibfnamefont {P.~S.}\ \bibnamefont
  {Swain}}\ and\ \bibinfo {author} {\bibfnamefont {D.}~\bibnamefont
  {Andelman}},\ }\href {\doibase 10.1103/PhysRevE.63.051911} {\bibfield
  {journal} {\bibinfo  {journal} {Phys. Rev. E}\ }\textbf {\bibinfo {volume}
  {63}},\ \bibinfo {pages} {051911} (\bibinfo {year} {2001})}\BibitemShut
  {NoStop}%
\bibitem [{\citenamefont {Zuckerman}\ and\ \citenamefont
  {Bruinsma}(1995)}]{Zuckerman1995}%
  \BibitemOpen
  \bibfield  {author} {\bibinfo {author} {\bibfnamefont {D.}~\bibnamefont
  {Zuckerman}}\ and\ \bibinfo {author} {\bibfnamefont {R.}~\bibnamefont
  {Bruinsma}},\ }\href {\doibase 10.1103/PhysRevLett.74.3900} {\bibfield
  {journal} {\bibinfo  {journal} {Phys. Rev. Lett.}\ }\textbf {\bibinfo
  {volume} {74}},\ \bibinfo {pages} {3900} (\bibinfo {year}
  {1995})}\BibitemShut {NoStop}%
\bibitem [{\citenamefont {Lipowsky}(1996)}]{Lipowsky1996}%
  \BibitemOpen
  \bibfield  {author} {\bibinfo {author} {\bibfnamefont {R.}~\bibnamefont
  {Lipowsky}},\ }\href {\doibase 10.1103/PhysRevLett.77.1652} {\bibfield
  {journal} {\bibinfo  {journal} {Phys. Rev. Lett.}\ }\textbf {\bibinfo
  {volume} {77}},\ \bibinfo {pages} {1652} (\bibinfo {year}
  {1996})}\BibitemShut {NoStop}%
\bibitem [{\citenamefont {Speck}\ and\ \citenamefont {Vink}(2012)}]{Speck2012}%
  \BibitemOpen
  \bibfield  {author} {\bibinfo {author} {\bibfnamefont {T.}~\bibnamefont
  {Speck}}\ and\ \bibinfo {author} {\bibfnamefont {R.~L.~C.}\ \bibnamefont
  {Vink}},\ }\href {\doibase 10.1103/PhysRevE.86.031923} {\bibfield  {journal}
  {\bibinfo  {journal} {Phys. Rev. E}\ }\textbf {\bibinfo {volume} {86}},\
  \bibinfo {pages} {031923} (\bibinfo {year} {2012})}\BibitemShut {NoStop}%
\bibitem [{\citenamefont {Hemmerle}\ \emph {et~al.}(2012)\citenamefont
  {Hemmerle}, \citenamefont {Malaquin}, \citenamefont {Charitat}, \citenamefont
  {Lecuyer}, \citenamefont {Fragneto},\ and\ \citenamefont
  {Daillant}}]{Hemmerle2013}%
  \BibitemOpen
  \bibfield  {author} {\bibinfo {author} {\bibfnamefont {A.}~\bibnamefont
  {Hemmerle}}, \bibinfo {author} {\bibfnamefont {L.}~\bibnamefont {Malaquin}},
  \bibinfo {author} {\bibfnamefont {T.}~\bibnamefont {Charitat}}, \bibinfo
  {author} {\bibfnamefont {S.}~\bibnamefont {Lecuyer}}, \bibinfo {author}
  {\bibfnamefont {G.}~\bibnamefont {Fragneto}}, \ and\ \bibinfo {author}
  {\bibfnamefont {J.}~\bibnamefont {Daillant}},\ }\href {\doibase
  10.1073/pnas.1211669109} {\ \textbf {\bibinfo {volume} {109}},\ \bibinfo
  {pages} {19938} (\bibinfo {year} {2012})}\BibitemShut {NoStop}%
\bibitem [{\citenamefont {Helfrich}(1978)}]{Helfrich1978}%
  \BibitemOpen
  \bibfield  {author} {\bibinfo {author} {\bibfnamefont {W.}~\bibnamefont
  {Helfrich}},\ }\href@noop {} {\bibfield  {journal} {\bibinfo  {journal} {Z.
  Naturforsch. A}\ }\textbf {\bibinfo {volume} {33}},\ \bibinfo {pages} {305}
  (\bibinfo {year} {1978})}\BibitemShut {NoStop}%
\bibitem [{\citenamefont {Freund}(2013)}]{Freund2013}%
  \BibitemOpen
  \bibfield  {author} {\bibinfo {author} {\bibfnamefont {L.~B.}\ \bibnamefont
  {Freund}},\ }\href {\doibase 10.1073/pnas.1220968110} {\ \textbf {\bibinfo
  {volume} {110}},\ \bibinfo {pages} {2047} (\bibinfo {year}
  {2013})}\BibitemShut {NoStop}%
\bibitem [{\citenamefont {Auth}\ and\ \citenamefont
  {Gompper}(2013)}]{Auth2013}%
  \BibitemOpen
  \bibfield  {author} {\bibinfo {author} {\bibfnamefont {T.}~\bibnamefont
  {Auth}}\ and\ \bibinfo {author} {\bibfnamefont {G.}~\bibnamefont {Gompper}},\
  }\href {\doibase 10.1103/PhysRevE.88.010701} {\bibfield  {journal} {\bibinfo
  {journal} {Phys. Rev. E}\ }\textbf {\bibinfo {volume} {88}},\ \bibinfo
  {pages} {010701} (\bibinfo {year} {2013})}\BibitemShut {NoStop}%
\bibitem [{\citenamefont {Weikl}\ \emph {et~al.}(2009)\citenamefont {Weikl},
  \citenamefont {Asfaw}, \citenamefont {Krobath}, \citenamefont {Rozycki},\
  and\ \citenamefont {Lipowsky}}]{Weikl2009}%
  \BibitemOpen
  \bibfield  {author} {\bibinfo {author} {\bibfnamefont {T.~R.}\ \bibnamefont
  {Weikl}}, \bibinfo {author} {\bibfnamefont {M.}~\bibnamefont {Asfaw}},
  \bibinfo {author} {\bibfnamefont {H.}~\bibnamefont {Krobath}}, \bibinfo
  {author} {\bibfnamefont {B.}~\bibnamefont {Rozycki}}, \ and\ \bibinfo
  {author} {\bibfnamefont {R.}~\bibnamefont {Lipowsky}},\ }\href {\doibase
  10.1039/B902017A} {\bibfield  {journal} {\bibinfo  {journal} {Soft Matter}\
  }\textbf {\bibinfo {volume} {5}},\ \bibinfo {pages} {3213} (\bibinfo {year}
  {2009})}\BibitemShut {NoStop}%
\bibitem [{\citenamefont {Bruinsma}\ \emph {et~al.}(2000)\citenamefont
  {Bruinsma}, \citenamefont {Behrisch},\ and\ \citenamefont
  {Sackmann}}]{Bruinsma2000}%
  \BibitemOpen
  \bibfield  {author} {\bibinfo {author} {\bibfnamefont {R.}~\bibnamefont
  {Bruinsma}}, \bibinfo {author} {\bibfnamefont {A.}~\bibnamefont {Behrisch}},
  \ and\ \bibinfo {author} {\bibfnamefont {E.}~\bibnamefont {Sackmann}},\
  }\href {\doibase 10.1103/PhysRevE.61.4253} {\bibfield  {journal} {\bibinfo
  {journal} {Phys. Rev. E}\ }\textbf {\bibinfo {volume} {61}},\ \bibinfo
  {pages} {4253} (\bibinfo {year} {2000})}\BibitemShut {NoStop}%
\bibitem [{\citenamefont {Sengupta}\ and\ \citenamefont
  {Limozin}(2010)}]{Sengupta2010}%
  \BibitemOpen
  \bibfield  {author} {\bibinfo {author} {\bibfnamefont {K.}~\bibnamefont
  {Sengupta}}\ and\ \bibinfo {author} {\bibfnamefont {L.}~\bibnamefont
  {Limozin}},\ }\href {\doibase 10.1103/PhysRevLett.104.088101} {\bibfield
  {journal} {\bibinfo  {journal} {Phys. Rev. Lett.}\ }\textbf {\bibinfo
  {volume} {104}},\ \bibinfo {pages} {088101} (\bibinfo {year}
  {2010})}\BibitemShut {NoStop}%
\bibitem [{\citenamefont {Canham}(1970)}]{Canham1970}%
  \BibitemOpen
  \bibfield  {author} {\bibinfo {author} {\bibfnamefont {P.~B.}\ \bibnamefont
  {Canham}},\ }\href@noop {} {\bibfield  {journal} {\bibinfo  {journal} {J.
  Theor. Biol}\ }\textbf {\bibinfo {volume} {26}},\ \bibinfo {pages} {61}
  (\bibinfo {year} {1970})}\BibitemShut {NoStop}%
\bibitem [{\citenamefont {Helfrich}(1973)}]{Helfrich1973}%
  \BibitemOpen
  \bibfield  {author} {\bibinfo {author} {\bibfnamefont {W.}~\bibnamefont
  {Helfrich}},\ }\href@noop {} {\bibfield  {journal} {\bibinfo  {journal} {Z.
  Naturforsch.}\ }\textbf {\bibinfo {volume} {28c}},\ \bibinfo {pages} {693}
  (\bibinfo {year} {1973})}\BibitemShut {NoStop}%
\bibitem [{\citenamefont {Lorz}\ \emph {et~al.}(2007)\citenamefont {Lorz},
  \citenamefont {Smith}, \citenamefont {Gege},\ and\ \citenamefont
  {Sackmann}}]{Lorz2007}%
  \BibitemOpen
  \bibfield  {author} {\bibinfo {author} {\bibfnamefont {B.~G.}\ \bibnamefont
  {Lorz}}, \bibinfo {author} {\bibfnamefont {A.-S.}\ \bibnamefont {Smith}},
  \bibinfo {author} {\bibfnamefont {C.}~\bibnamefont {Gege}}, \ and\ \bibinfo
  {author} {\bibfnamefont {E.}~\bibnamefont {Sackmann}},\ }\href {\doibase
  10.1021/la701824q} {\bibfield  {journal} {\bibinfo  {journal} {Langmuir}\
  }\textbf {\bibinfo {volume} {23}},\ \bibinfo {pages} {12293} (\bibinfo {year}
  {2007})}\BibitemShut {NoStop}%
\bibitem [{\citenamefont {Reister-Gottfried}\ \emph {et~al.}(2008)\citenamefont
  {Reister-Gottfried}, \citenamefont {Sengupta}, \citenamefont {Lorz},
  \citenamefont {Sackmann}, \citenamefont {Seifert},\ and\ \citenamefont
  {Smith}}]{ReisterGottfried2008}%
  \BibitemOpen
  \bibfield  {author} {\bibinfo {author} {\bibfnamefont {E.}~\bibnamefont
  {Reister-Gottfried}}, \bibinfo {author} {\bibfnamefont {K.}~\bibnamefont
  {Sengupta}}, \bibinfo {author} {\bibfnamefont {B.}~\bibnamefont {Lorz}},
  \bibinfo {author} {\bibfnamefont {E.}~\bibnamefont {Sackmann}}, \bibinfo
  {author} {\bibfnamefont {U.}~\bibnamefont {Seifert}}, \ and\ \bibinfo
  {author} {\bibfnamefont {A.-S. c.~v.}\ \bibnamefont {Smith}},\ }\href
  {\doibase 10.1103/PhysRevLett.101.208103} {\bibfield  {journal} {\bibinfo
  {journal} {Phys. Rev. Lett.}\ }\textbf {\bibinfo {volume} {101}},\ \bibinfo
  {pages} {208103} (\bibinfo {year} {2008})}\BibitemShut {NoStop}%
\bibitem [{\citenamefont {Bihr}\ \emph {et~al.}(2012)\citenamefont {Bihr},
  \citenamefont {Seifert},\ and\ \citenamefont {Smith}}]{Bihr2012}%
  \BibitemOpen
  \bibfield  {author} {\bibinfo {author} {\bibfnamefont {T.}~\bibnamefont
  {Bihr}}, \bibinfo {author} {\bibfnamefont {U.}~\bibnamefont {Seifert}}, \
  and\ \bibinfo {author} {\bibfnamefont {A.-S. c.~v.}\ \bibnamefont {Smith}},\
  }\href {\doibase 10.1103/PhysRevLett.109.258101} {\bibfield  {journal}
  {\bibinfo  {journal} {Phys. Rev. Lett.}\ }\textbf {\bibinfo {volume} {109}},\
  \bibinfo {pages} {258101} (\bibinfo {year} {2012})}\BibitemShut {NoStop}%
\bibitem [{\citenamefont {Kusumi}\ \emph {et~al.}(2005)\citenamefont {Kusumi},
  \citenamefont {Nakada}, \citenamefont {Ritchie}, \citenamefont {Murase},
  \citenamefont {Suzuki}, \citenamefont {Murakoshi}, \citenamefont {Kasai},
  \citenamefont {Kondo},\ and\ \citenamefont {Fujiwara}}]{Kusumi2005}%
  \BibitemOpen
  \bibfield  {author} {\bibinfo {author} {\bibfnamefont {A.}~\bibnamefont
  {Kusumi}}, \bibinfo {author} {\bibfnamefont {C.}~\bibnamefont {Nakada}},
  \bibinfo {author} {\bibfnamefont {K.}~\bibnamefont {Ritchie}}, \bibinfo
  {author} {\bibfnamefont {K.}~\bibnamefont {Murase}}, \bibinfo {author}
  {\bibfnamefont {K.}~\bibnamefont {Suzuki}}, \bibinfo {author} {\bibfnamefont
  {H.}~\bibnamefont {Murakoshi}}, \bibinfo {author} {\bibfnamefont {R.~S.}\
  \bibnamefont {Kasai}}, \bibinfo {author} {\bibfnamefont {J.}~\bibnamefont
  {Kondo}}, \ and\ \bibinfo {author} {\bibfnamefont {T.}~\bibnamefont
  {Fujiwara}},\ }\href {\doibase 10.1146/annurev.biophys.34.040204.144637}
  {\bibfield  {journal} {\bibinfo  {journal} {Annual Review of Biophysics and
  Biomolecular Structure}\ }\textbf {\bibinfo {volume} {34}},\ \bibinfo {pages}
  {351} (\bibinfo {year} {2005})},\ \bibinfo {note} {pMID:
  15869394}\BibitemShut {NoStop}%
\bibitem [{\citenamefont {Delano\"e-Ayari}\ \emph {et~al.}(2004)\citenamefont
  {Delano\"e-Ayari}, \citenamefont {Lenz}, \citenamefont {Brevier},
  \citenamefont {Weidenhaupt}, \citenamefont {Vallade}, \citenamefont {Gulino},
  \citenamefont {Joanny},\ and\ \citenamefont {Riveline}}]{DelanoeAyari2004}%
  \BibitemOpen
  \bibfield  {author} {\bibinfo {author} {\bibfnamefont {H.}~\bibnamefont
  {Delano\"e-Ayari}}, \bibinfo {author} {\bibfnamefont {P.}~\bibnamefont
  {Lenz}}, \bibinfo {author} {\bibfnamefont {J.}~\bibnamefont {Brevier}},
  \bibinfo {author} {\bibfnamefont {M.}~\bibnamefont {Weidenhaupt}}, \bibinfo
  {author} {\bibfnamefont {M.}~\bibnamefont {Vallade}}, \bibinfo {author}
  {\bibfnamefont {D.}~\bibnamefont {Gulino}}, \bibinfo {author} {\bibfnamefont
  {J.~F.}\ \bibnamefont {Joanny}}, \ and\ \bibinfo {author} {\bibfnamefont
  {D.}~\bibnamefont {Riveline}},\ }\href {\doibase
  10.1103/PhysRevLett.93.108102} {\bibfield  {journal} {\bibinfo  {journal}
  {Phys. Rev. Lett.}\ }\textbf {\bibinfo {volume} {93}},\ \bibinfo {pages}
  {108102} (\bibinfo {year} {2004})}\BibitemShut {NoStop}%
\bibitem [{\citenamefont {Mueller}\ and\ \citenamefont
  {Mueller-Plathe}(2009)}]{Mueller2009}%
  \BibitemOpen
  \bibfield  {author} {\bibinfo {author} {\bibfnamefont {T.~J.}\ \bibnamefont
  {Mueller}}\ and\ \bibinfo {author} {\bibfnamefont {F.}~\bibnamefont
  {Mueller-Plathe}},\ }\href {\doibase 10.1002/cphc.200900156} {\bibfield
  {journal} {\bibinfo  {journal} {ChemPhysChem}\ }\textbf {\bibinfo {volume}
  {10}},\ \bibinfo {pages} {2305} (\bibinfo {year} {2009})}\BibitemShut
  {NoStop}%
\bibitem [{\citenamefont {von Hansen}\ \emph {et~al.}(2013)\citenamefont {von
  Hansen}, \citenamefont {Gekle},\ and\ \citenamefont {Netz}}]{VonHansen2013}%
  \BibitemOpen
  \bibfield  {author} {\bibinfo {author} {\bibfnamefont {Y.}~\bibnamefont {von
  Hansen}}, \bibinfo {author} {\bibfnamefont {S.}~\bibnamefont {Gekle}}, \ and\
  \bibinfo {author} {\bibfnamefont {R.~R.}\ \bibnamefont {Netz}},\ }\href
  {\doibase 10.1103/PhysRevLett.111.118103} {\bibfield  {journal} {\bibinfo
  {journal} {Phys. Rev. Lett.}\ }\textbf {\bibinfo {volume} {111}},\ \bibinfo
  {pages} {118103} (\bibinfo {year} {2013})}\BibitemShut {NoStop}%
\bibitem [{\citenamefont {Oron}\ \emph {et~al.}(1997)\citenamefont {Oron},
  \citenamefont {Davis},\ and\ \citenamefont {Bankoff}}]{Oron1997}%
  \BibitemOpen
  \bibfield  {author} {\bibinfo {author} {\bibfnamefont {A.}~\bibnamefont
  {Oron}}, \bibinfo {author} {\bibfnamefont {S.~H.}\ \bibnamefont {Davis}}, \
  and\ \bibinfo {author} {\bibfnamefont {S.~G.}\ \bibnamefont {Bankoff}},\
  }\href {\doibase 10.1103/RevModPhys.69.931} {\bibfield  {journal} {\bibinfo
  {journal} {Rev. Mod. Phys.}\ }\textbf {\bibinfo {volume} {69}},\ \bibinfo
  {pages} {931} (\bibinfo {year} {1997})}\BibitemShut {NoStop}%
\bibitem [{\citenamefont {Happel}\ and\ \citenamefont
  {Brenner}(1983)}]{Happel1983}%
  \BibitemOpen
  \bibfield  {author} {\bibinfo {author} {\bibfnamefont {J.}~\bibnamefont
  {Happel}}\ and\ \bibinfo {author} {\bibfnamefont {H.}~\bibnamefont
  {Brenner}},\ }\href@noop {} {\emph {\bibinfo {title} {Low Reynolds Number
  Hydrodynamics}}}\ (\bibinfo  {publisher} {Kluwer},\ \bibinfo {year}
  {1983})\BibitemShut {NoStop}%
\bibitem [{\citenamefont {Peletier}\ and\ \citenamefont
  {Troy}(2001)}]{Peletier2001}%
  \BibitemOpen
  \bibfield  {author} {\bibinfo {author} {\bibfnamefont {L.}~\bibnamefont
  {Peletier}}\ and\ \bibinfo {author} {\bibfnamefont {W.}~\bibnamefont
  {Troy}},\ }\href@noop {} {\emph {\bibinfo {title} {Spatial Patterns}}}\
  (\bibinfo  {publisher} {Birkhauser Boston},\ \bibinfo {year}
  {2001})\BibitemShut {NoStop}%
\bibitem [{\citenamefont {Le~Goff}\ \emph {et~al.}(2014)\citenamefont
  {Le~Goff}, \citenamefont {Politi},\ and\ \citenamefont
  {Pierre-Louis}}]{LeGoff2014}%
  \BibitemOpen
  \bibfield  {author} {\bibinfo {author} {\bibfnamefont {T.}~\bibnamefont
  {Le~Goff}}, \bibinfo {author} {\bibfnamefont {P.}~\bibnamefont {Politi}}, \
  and\ \bibinfo {author} {\bibfnamefont {O.}~\bibnamefont {Pierre-Louis}},\
  }\href@noop {} {\bibfield  {journal} {\bibinfo  {journal} {unpublished}\ }
  (\bibinfo {year} {2014})}\BibitemShut {NoStop}%
\bibitem [{\citenamefont {Malaquin}\ \emph {et~al.}(2010)\citenamefont
  {Malaquin}, \citenamefont {T.~Charitat},\ and\ \citenamefont
  {Daillant}}]{Malaquin2010}%
  \BibitemOpen
  \bibfield  {author} {\bibinfo {author} {\bibfnamefont {L.}~\bibnamefont
  {Malaquin}}, \bibinfo {author} {\bibfnamefont {T.}~\bibnamefont
  {T.~Charitat}}, \ and\ \bibinfo {author} {\bibfnamefont {J.}~\bibnamefont
  {Daillant}},\ }\href@noop {} {\bibfield  {journal} {\bibinfo  {journal} {The
  European Physical J E}\ }\textbf {\bibinfo {volume} {31}},\ \bibinfo {pages}
  {285} (\bibinfo {year} {2010})}\BibitemShut {NoStop}%
\end{thebibliography}
\end{document}